\documentstyle[prl,aps,floats,twocolumn]{revtex}
\input{psfig.tex}
\begin{document}

\draft
\twocolumn[\hsize\textwidth\columnwidth\hsize\csname @twocolumnfalse\endcsname

\title{Measuring Dark Matter Power Spectrum from 
Cosmic Microwave Background }
\author{Uro\v s Seljak\cite{urosemail}\cite{presentaddress}}
\address{Max-Planck Institut fuer Astrophysik,
Karl-Schwartzschild-Str. 1,
85740 Garching beim Muenchen,
Germany }
\author{Matias Zaldarriaga\cite{matiasemail}}
\address{Institute for Advanced Studies, School of Natural Sciences, Princeton, NJ 08540}
\date{October 1998}
\maketitle

\begin{abstract}
We propose a method to extract the projected power spectrum of density 
perturbations from the distortions in the cosmic microwave background (CMB).
The distortions are imprinted onto the CMB by the gravitational lensing effect
and can be extracted using a combination of products of CMB derivatives. We 
show that future CMB experiments such as Planck will be able to extract the 
power spectrum with high statistical significance over two orders of magnitude
in angle.
The method proposed here traces dark matter directly to 
higher redshift (up to $z \sim 1100$) and larger scale 
(few Gpc) than 
any other currently known method. It also traces large scale structure in 
the linear regime, allowing simple interpretation in terms of cosmological 
models. 
By providing additional and complementary information to the one 
from the primary CMB analysis 
it will strengthen further the scientific return of future
CMB experiments. 
\end{abstract}
\pacs{PACS numbers: 98.80.Es,95.85.Bh,98.35.Ce,98.70.Vc  \hfill}
]
\def\bi#1{\bbox{#1}}
\def\gsim{\raise2.90pt\hbox{$\scriptstyle
>$} \hspace{-6.4pt}
\lower.5pt\hbox{$\scriptscriptstyle
\sim$}\; }
\def\lsim{\raise2.90pt\hbox{$\scriptstyle
<$} \hspace{-6pt}\lower.5pt\hbox{$\scriptscriptstyle\sim$}\; }
\newcommand{\tl}{\tilde}

The standard paradigm of modern cosmology 
is the expanding universe model, where the universe started with 
a hot big bang and has been expanding ever since. The universe
is predicted to be 
homogeneous and isotropic on large scales with initially small 
fluctuations superimposed on that background. These fluctuation 
have been amplified by 
gravity to become the present day large scale structure.
This paradigm, while successful in explaining all the observations, still
leaves many questions unanswered. Among these are the value
of cosmological parameters, such as the density of the various  
components, the nature of dark matter and the power spectrum of
initial perturbations. 

Observational efforts are necessary to provide the 
answers to the  above questions.
It has long been recognized that
the power spectrum of dark matter fluctuations 
would provide many of these answers, as it depends both on the 
initial power spectrum as well as on the rate of growth of perturbations, 
which are determined by the cosmological model and the nature of dark matter
(e.g. hot or cold).  
Its measurement has been the primary target of many 
observational surveys in the past and this trend will also continue
in the future. Most of these have used galaxy positions and redshifts
to map the universe.  
The ongoing surveys such as SDSS and 2dF 
hope to measure the power spectrum accurately 
on scales up to 
several hundred megaparsecs. 
However, there are a number of possible uncertainties that may 
make this goal more difficult than expected. 
In particular, the fact that galaxy formation
is such a poorly understood process makes 
the connection between mass and light uncertain. The relation 
between dark matter and galaxy density perturbation could be 
nonlinear, scale-dependent, non-local or stochastic even on 
very large scales \cite{blanton}. Another complication is the fact 
that except on very large scales, the surveys are measuring 
the clustering in the nonlinear regime, which makes the interpretation and 
statistical analysis of the data more complicated.

For these reasons
other, more direct, tracers of dark matter have been investigated, most 
notably velocity 
flows \cite{sw}, clustering of Ly-$\alpha$ forest \cite{croft} 
and weak lensing 
\cite{kaiser}. Each of these has uncertainties associated with them.
Weak lensing, although not yet detected, offers
perhaps the best chance of
measuring accurately the power spectrum of dark matter over a
large range of scales.
By measuring the distortions in the shapes of
background galaxies deformed by the mass distribution along the 
line of sight one can determine the clustering fluctuations
on scales from 1 up to 100 $h^{-1}$Mpc \cite{kaiser,js,wbm}. 
Weak lensing has the 
advantage of tracing the dark matter directly and therefore avoids
the uncertainties connected with galaxy clustering. Still, there are
uncertainties associated with weak lensing as well, 
the most important being the poorly known redshift distribution of 
background galaxies and their possible 
intrinsic alignment.

In this letter we propose a new method to measure the projected
dark matter power spectrum by using the weak lensing 
distortions of the  CMB \cite{bern}. 
We show that gravitational lensing 
induces characteristic distortions in the pattern of the CMB field,
which allows one to reconstruct the projected dark matter density 
field \cite{sz98}. The reconstructed projected mass density 
can be used to measure
the mass  power spectrum 
by averaging over independent CMB patches.
The method proposed here 
offers several advantages over
the other methods discussed above. It traces the
dark matter directly  
and does not depend on the assumptions of how light traces mass.
Unlike weak lensing of galaxies it
does not suffer from the possible 
intrinsic alignment of background galaxies or their 
uncertain redshift distribution. 
It is able to recover the power spectrum
over a large range of scales, roughly between 10 Mpc and 1 Gpc. As such it is
sensitive to scales larger than any other survey, 
including the 
SDSS and 2dF. Furthermore,
the observed field is a projection of density up to $z \sim 1100$ and one 
is therefore tracing dark matter to higher 
redshifts than with any other method. The last two features also imply that with 
the proposed method one 
is tracing the perturbations predominantly in the linear regime and  
no nonlinear corrections are necessary 
down to the smallest scales that are still 
observable. 
In many ways the method proposed here
offers the same advantages that make the CMB anisotropies 
such a powerful test 
of cosmological models, except that it is providing complementary 
information on the projected dark matter power spectrum. 
Clearly, combining the standard CMB analysis with the present
one will further enhance the power of CMB to constrain the cosmological models
\cite{eth} and provide additional motivation for future CMB experiments.

To provide a clear physical interpretation we will
only discuss the small scale limit of the method in this letter 
(see \cite{sz98,isw} for all-sky generalization). 
The observed CMB temperature in the direction $\bi \theta$ is
$T({\bi \theta})$ and
equals the (unobservable) temperature at the last scattering surface
in a different direction, $
\tl T({\bi \theta}+\delta {\bi \theta})$, where
$\delta {\bi \theta}$ is the angular excursion of the photon as it
propagates from the last scattering
surface to us.
In terms of Fourier components we have
$T({\bi \theta})
         =(2\pi)^{-2}\int d^2{\bi l}\
e^{i{\bi l}\cdot (\bi{\theta}+ \delta
{\bi \theta})}\ \tl T({\bi l})
$.
To extract the information on the deflection field
$\delta{\bi \theta}$ we consider derivatives
of the CMB temperature. If the CMB is an isotropic and
homogeneous Gaussian random
field then different partial derivatives are
statistically equivalent and their spatial properties
are independent of position. Lensing will distort
these two properties of the derivatives.
The derivatives of the temperature field
are to lowest order, 
\begin{equation}
T_a({\bi \theta})\equiv {\partial \tl T
\over \partial \theta_a} (\bi{\theta}+\delta{\bi
\theta}) \nonumber \\
= (\delta_{ab}+\Phi_{ab}) \tl T_b(\bi{\theta}+\delta{\bi\theta}),
\label{dert}
\end{equation}
where $\Phi_{ab}= {\partial \delta \theta_a \over \partial \theta_b}$
is the symmetric shear tensor.
The components of the shear tensor can be written in terms of the
convergence $\kappa=-(\Phi_{aa}+\Phi_{bb})/2$ and the two shear components 
$\gamma_1=-(\Phi_{aa}-\Phi_{bb})/2$ and
$\gamma_2=-\Phi_{ab}$.

Next we consider the quadratic combinations of the derivatives 
and express
them in terms of the unlensed field to lowest order in the shear tensor,
\begin{eqnarray}
{\cal S} & \equiv & 1-\sigma_{\cal S}^{-1}\left[T_x^2+T_y^2\right]({\bi \theta}) 
\nonumber \\
       & =&1-\sigma_{\cal S}^{-1}\left[(1+\Phi_{xx}+\Phi_{yy})\tl {\cal S}+(\Phi_{xx}-\Phi_{yy})
        \tl {\cal Q} +
        2 \Phi_{xy} \tl{\cal U}\right] \nonumber \\
{\cal Q}&\equiv&-\sigma_{\cal S}^{-1}\left[T_x^2-T_y^2\right]({\bi \theta}) 
 \nonumber \\      & =&-\sigma_{\cal S}^{-1}\left[(1+\Phi_{xx}+\Phi_{yy})\tl {\cal Q}+(\Phi_{xx}-\Phi_{yy}) \tl {\cal S}\right]
        \nonumber \\
{\cal U}&\equiv&-2 \sigma_{\cal S}^{-1}\left[T_x T_y\right]({\bi \theta}) 
 \nonumber \\   &=&- 2 \sigma_{\cal S}^{-1}
\left[(1+\Phi_{xx}+\Phi_{yy})\tl {\cal U}+\Phi_{xy} \tl {\cal S}\right] ,
\label{derivs}
\end{eqnarray}
where $\sigma_{\cal S}$ was defined so that in the absence of lensing
$\langle {\cal S} \rangle=\langle {\cal Q} \rangle=\langle {\cal U} \rangle=0$.
In the presence of lensing it follows from above 
$\langle {\cal S} \rangle=
2 \kappa$, $\langle {\cal Q} \rangle=2\gamma_1$, $\langle {\cal U} \rangle=2\gamma_2$.
Physically, $\kappa$ stretches the image and makes the derivatives smaller. Similarly,
shear produces anisotropy in the derivatives and can be extracted by considering
the particular combination of derivatives defined above. From ${\cal Q}$ and 
${\cal U}$ we can form two rotationally invariant quantities in Fourier space
\begin{eqnarray}
{\cal E}(\bi{l})&\equiv&{\cal Q}(\bi{l})\cos(2\phi_{\bi{l}})+{\cal U}(\bi{l})\sin(2\phi_{\bi{l}})
\nonumber \nonumber \\
{\cal B}(\bi{l})&\equiv&{\cal Q}(\bi{l})\sin(2\phi_{\bi{l}})-{\cal U}(\bi{l})\cos(2\phi_{\bi{l}}).
\end{eqnarray} 
The average of the scalar field $\cal E$ is 
$\langle {\cal E} \rangle =2 \kappa$. The
average of the pseudo-scalar field  
$\langle {\cal B} \rangle$ vanishes in the large scale limit
because gravitational 
potential from which shear is generated is invariant under the parity 
transformation. The convergence
$\kappa$ can be reconstructed 
in two independent ways, either from $\cal S$ or $\cal E$, 
while $\cal B$ can serve as 
a check for possible systematics.

The convergence $\kappa$ can be written 
as a projection of gravitational potential 
$\kappa= \int_0^{\chi_0} g(\chi,\chi_0)\nabla^2\phi(\chi) d \chi$ \cite{kaiser,js}.
Here $\chi_0$ is the comoving radial coordinate at recombination and
$g$ is the radial window, defined as $g(\chi,\chi_0) = 
{r(\chi)r(\chi_0-\chi) \over r(\chi_0) }$.
Here $r(\chi)$ is the comoving angular diameter
distance, defined as $K^{-1/2}\sin K^{1/2}\chi$,
$\chi$, $(-K)^{-1/2}\sinh (-K)^{1/2}\chi$ for $K>0$, $K=0$, $K<0$,
respectively. The curvature $K$
can be expressed using the present density
parameter $\Omega_0$
and the present
Hubble parameter $H_0$ as $K=(\Omega_0-1)H_0^2$.
In general $\Omega_0$ consists both of matter contribution 
$\Omega_m$ and cosmological constant term $\Omega_{\Lambda}$.

\begin{figure}[htbp]
\centerline{\psfig{file=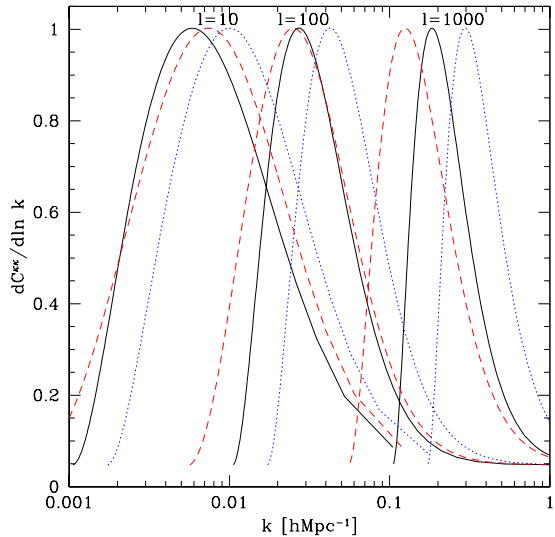,width=3.0in}}
\caption{
Logarithmic contribution to $C^{\kappa \kappa}_l$
as a function of $k$ for $l=10$, 100, 1000 (the normalization is
arbitrary).
The models are flat CDM model (dotted), open CDM model with $\Omega_m=0.3$
(dashed) and cosmological constant model with with $\Omega_m=0.3$ (solid). All 
the models have $\Gamma=\Omega_m h=0.21$.}
\label{fig1}
\end{figure}

The angular 
power spectrum of convergence $C_l^{\kappa \kappa}$ has ensemble  
average \cite{js}
\begin{equation}
C^{\kappa \kappa}_l={9 \over 4} \Omega_m^2\,
\int_0^{\chi_0}\ {g^2(\chi,\chi_0) \over a^2(\chi)r^2(\chi)}
\ P_\delta\left(k={l \over r(\chi)},\chi\right)d\chi,
\label{cl}
\end{equation}
where $a(\chi)$ is the expansion factor and 
we used Poisson's equation to express the convergence directly 
in terms of time dependent matter power spectrum $P_\delta(k,\tau)$. 
Figure \ref{fig1} shows the contribution 
of the 3-d power spectrum $P_{\delta}(k)$ to $C_l^{\kappa \kappa}$
as a function of $k$ for different $l$. It is a relatively broad 
function of $k$. Different cosmological models 
give somewhat different correspondence between $l$ and $k$, 
but in general $l=10$ 
probes scales around $\lambda=2\pi/k=1h^{-1}$Gpc, while $l=1000$
contribution probes scales around $\lambda=30h^{-1}$Mpc. We will see below
that this spans the observable range of Planck satellite.  

\begin{figure}
\centerline{\psfig{file=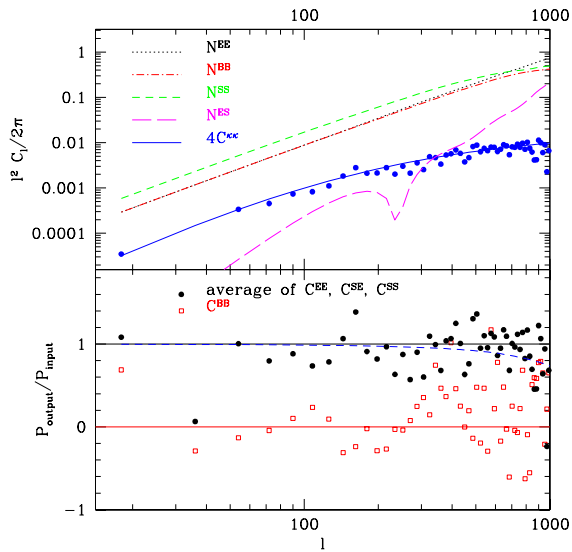,width=3.0in}}
\caption{Top: power spectra for noise $N^{\cal S \cal S}_l$ (short dashed), 
$N^{\cal E\cal E}_l$ (dotted), $N^{\cal E \cal S}_l$ (long-dashed) and 
$N^{\cal BB}_l$ (dash-dotted)
for Planck, using cosmological 
constant model with $\Omega_m=0.3$ and $\sigma_8=1$. Also shown is power 
spectrum of convergence $ 4C^{\kappa \kappa}_l$ (solid) for the same model, 
together with its reconstruction from a Monte Carlo simulation. 
Bottom: ratio of output to input power spectrum is plotted for the 
averaged $C^{\kappa \kappa}_l$ reconstruction and for the $C^{\cal BB}_l$
reconstruction. 
Also shown is the window $W_l$ (dashed) for Planck satellite characteristics.
}
\label{fig2}
\end{figure}

From $\cal S$ and $\cal E$ one can form three different angular 
power spectra, 
$C_l^{\cal SS}$, $C_l^{\cal EE}$ and $C_l^{\cal SE}$, defined as 
$C^{\cal WW'}_l= [{\cal W}({\bi l})^*{\cal W}'({\bi l'})+
{\cal W}'({\bi l'})^*{\cal W}({\bi l})]/2\ \delta_{ll'}$. 
Their expectation value is  
$\langle C_l^{{\cal WW}'} 
\rangle=4W_lC^{\kappa \kappa}_l+N^{\cal WW'}_l$.  
Here $N^{\cal WW'}_l$ are the 
noise power spectra for $\cal S$, $\cal E$ or their cross-term, while $W_l$ is the window 
function describing the signal degradation because of finite angular resolution and 
detector noise \cite{sz98}.
The window changes smoothly from unity at low $l$ to zero at high $l$. 
For MAP the transition occurs at $l \sim 300$,
while for Planck the window is close to unity up to $l \sim 1000$. The noise power spectra
have contribution from intrinsic CMB 
fluctuations and detector noise. In the absence
of detector noise it is the correlation length of the CMB $\xi$ that governs the 
amplitude of noise on large scales. Each patch of the sky of size 
$\xi^2$
contributes one independent measurement with 
a variance of order unity. By averaging over many independent patches we can 
reduce the noise on large scales. 
Detector noise and beam smoothing degrade this
and the CMB field has to be smoothed at the angular scale where the 
detector noise exceeds
the CMB signal \cite{sz98}. This increases 
the CMB correlation length and the amplitude of noise power spectrum.
Beam and noise 
characteristics of Planck ($0.12^{\circ}$ FWHM and $N^{TT}=(0.01\mu K)^2$) make
the influence of detector noise negligible, as there is very little CMB 
power on small scales which are not accessible by Planck. 
The noise spectra for $\cal SS$, 
$\cal EE$, $\cal BB$ and $\cal SE$ 
are shown in top of figure \ref{fig2}. On large scales 
$N^{\cal EE} \approx N^{\cal BB} \approx N^{\cal SS}/2 \gg N^{\cal SE} $.
Also shown is the typical convergence power spectrum 
$4C_l^{\kappa \kappa}$, which is 
significantly below Planck noise for $\cal S$ or $\cal E$. This  
shows that the 
reconstruction can only be achieved in a statistical sense by averaging over independent 
multipole moments.
For MAP with $0.21^{\circ}$ FWHM and $N^{TT}=(0.11\mu K)^2$ 
the noise power spectra are a factor of 5 higher than for Planck \cite{sz98}. 
From the measured $C_l^{{\cal WW}'}$ 
we subtract the noise power spectra $N^{\cal WW'}_l$
and are left with an estimate of $C_l \equiv 4W_lC_l^{\kappa \kappa}$.

To combine the three estimates of $C_l$
obtained from  $\cal SS$, 
$\cal EE$ and $\cal SE$ correlations
in an optimal way, we consider their covariance matrix. 
If the CMB noise can be considered
Gaussian then the covariance matrix for the estimated power spectra can be
expressed in terms of the noise power spectra. 
We have verified using Monte 
Carlo simulations that the 
Gaussian approximation is an excellent one in the large scale 
limit \cite{sz98}, despite the fact that the underlying fields are not Gaussian (they
are constructed using squares of the temperature field). This is
a consequence
of the central limit theorem, as there
are many uncorrelated patches that contribute to the low $l$ modes. 
The diagonal 
terms of the covariance matrix are given by
${\rm Cov}[(C^{\cal WW})^2]={2\over 2l+1} (N^{\cal W W}_l)^2$ and 
${\rm Cov}[(C^{\cal SE})^2]={1\over 2l+1} [N^{\cal SS}_lN^{\cal EE}_l+(N^{\cal SE}_l)^2]$.
The off-diagonal elements can be ignored compared to the 
diagonal terms at low
$l$ because $N^{\cal SE}_l \ll N^{\cal SS}_l,
N^{\cal EE}_l$. The covariance 
matrix is then diagonal and 
we can use simple inverse noise variance weighting to find the 
best combination 
\begin{equation}
{\hat C}_l=\sigma_{C_l}^{-2}\sum_{{\cal WW}'={\cal SS},{\cal EE}, {\cal SE}}{C_l^{\cal WW'}-N_l^{\cal WW'} \over {\rm Cov}(C^{\cal WW'})^2},
\label{waverage}
\end{equation}
where 
$\sigma_{C_l}^{-2}={\rm Cov}^{-1}[(C^{\cal SS})^{2}]+{\rm Cov}^{-1}
[(C^{\cal EE})^2]+
{\rm Cov}^{-1}[(C^{\cal SE})^2]$ is the variance of $\hat C_l$.

The reconstructed average multiplied with $W_l^{-1}$ 
is plotted in figure \ref{fig2} for the Planck experiment using one Monte 
Carlo realization of the sky. We find that the 
input power spectrum is recovered 
up to $l \sim 1000$. The ratio between the two, shown in bottom of 
figure \ref{fig2},
is consistent with unity over this range, while the corresponding ratio for $\cal B$ is consistent with 0.  
There is no significant bias in $\cal S$, $\cal E$ or $\cal B$. 

Note that even for Planck 
noise is larger than the signal by a large factor, 
$(S/N)_l^{-1}=N^{\cal EE}_l/4W_lC^{\kappa \kappa}_l \sim 10$
around $l \sim 20$ for this model. Therefore we need about 
$2(S/N)_l^{-2} \sim 200$ modes to reach $S/N=1$ on the power 
spectrum and this in only possible if $l > (2/f_{\rm sky})^{1/2}(S/N)_l 
\sim 14f_{\rm sky}^{-1/2}$, where 
$f_{\rm sky}$ is the sky fraction that is being observed. 
This means that we cannot successfully recover the power spectrum for $l<14$
with $S/N>1$,
so the small scale analysis in this paper is adequate.
However, the information from  low $l$ modes can still be useful. There 
may be models which predict a large increase in power on very large
(Gpc) scales. 
For such models noise may
be below the signal and should give a detectable signal. Also,
low $l$ modes can be cross-correlated with other maps to enhance the 
signal to noise. One example is the CMB itself, where
a positive detection would be a signature 
of a time dependent gravitational potential \cite{isw,dg,ct}.

For MAP the reconstruction above fails to give a positive detection. 
A more detailed statistical
analysis is necessary to asses the signal to noise when all the information is 
combined. 
To asses the overall signal to noise we combine the information from all 
the 
estimates $X=\sum_{l} \alpha_l \hat{C}_l$, such that 
$S/N=\langle X \rangle /\langle X^2 \rangle^{1/2}=\sum_{l} \alpha_l 
\hat{C}_l/(\sum_{l}\alpha_l^2 \sigma_{C_l}^2)^{1/2}$ is maximized. 
This is achieved by 
$\alpha_l=C_l/\sigma_{C_l}^2$ \cite{isw}. The resulting 
signal to noise is 
\begin{equation}
{ S \over N}= \left[ f_{sky}
\sum_l (2l+1) {16 W_l^2 (C^{\kappa \kappa}_l)^2 \over \sigma_{C_l}^2
} \right]^{1/2}.
\end{equation}
For MAP this gives $S/N=3$ for 
$\Omega_m=0.3$, $\sigma_8=1$ model with cosmological 
constant. Other viable models give comparable numbers. 
MAP detection will therefore only be 
marginal, unless the power spectrum on large scales turns out to be much larger 
than expected. For Planck the viable models give $S/N$ between 15-25, confirming
the conclusion above that Planck will be able to extract the power spectrum
with high statistical significance.  

We have also investigated possible sources of systematics.
One is the fact that we have to subtract the CMB noise, which depends on 
the detailed 
knowledge of the unlensed CMB power spectrum, 
while only the lensed power spectrum is known with finite 
accuracy. We have found both effects  
to be below 0.5\% of the noise
amplitude of Planck, 
so they will not significantly affect the power spectrum reconstruction
for Planck, but will be an additional source of error for MAP
\cite{sz98}.
Another possible
uncertainty is the assumption of the Gaussian 
statistics, which allows us to estimate the noise and subtract it from the 
measurements. It should be noted that 
even if there is a non Gaussian component in the CMB (for example due
to foregrounds and secondary processes), 
the Gaussian approximation may still be valid 
on large scales because of the central limit theorem. The accuracy of
subtraction 
is more critical for $\hat{C}^{\cal SS}$ and $\hat{C}^{\cal EE}$ than 
for $\hat{C}^{\cal SE}$, where the noise
is lower than the signal for $l< 300$ (figure \ref{fig2}).
If only cross-correlation
is used to estimate $C^{\kappa \kappa}$ then the signal to noise is 
becomes roughly 2/3 of our previous  estimates 
and would still be clearly detectable with Planck. 
Finally, there are many 
consistency checks that can be applied to the reconstruction, 
such as comparing the
three estimators which should all be consistent 
with each other and verifying that 
$\hat{C}^{\cal BB}$ is consistent with 0. 
For low $l$ one may only use 
$\hat{N}^{\cal BB}$ to subtract the noise
directly, as in this limit we have  $N^{\cal BB}=N^{\cal EE}
=N^{\cal SS}/2$. 
To summarize, future CMB experiments  
have the potential of providing another important map of the 
universe, allowing one to trace the 
large scale distribution of dark matter 
to higher redshifts and larger scales than any other method.  

\smallskip
U.S. and M.Z. would like to thank Observatoire de Strasbourg and MPA, 
Garching, respectively, for 
hospitality during the visits.  
M.Z. is supported by NASA through Hubble Fellowship grant
HF-01116.01-98A from STScI,
operated by AURA, Inc. under NASA contract NAS5-26555.

\vfil\eject
 
\end{document}